\documentstyle[12pt]{article}
\newcommand{\affa}{Theory Group, Tata Institute of Fundamental Research,
   Homi Bhabha Road, Bombay 400005, India, Email--- saumen@theory.tifr.res.in}
\newcommand{\affb}{Theory Group, Tata Institute of Fundamental Research,
   Homi Bhabha Road, Bombay 400005, India, Email--- sgupta@theory.tifr.res.in}
\newcommand{\affc}{Theory Group, Tata Institute of Fundamental Research,
   Navrangpura, Ahmedabad 380009, India, Email--- ravi@prl.ernet.in}

\begin{document}
\begin{titlepage}
\begin{flushright}\vbox{\begin{tabular}{c}
           TIFR/TH/96-51\\
           September, 1996\\
           hep-ph/9609407
\end{tabular}}\end{flushright}
\begin{center}
   {\large \bf  Collinear Divergences at One-loop Order\\
                for External Particles in a Heat-bath.}
\end{center}
\bigskip
\begin{center}
   {Saumen Datta\footnote\affa,
    Sourendu Gupta\footnote\affb,
    V.\ Ravindran\footnote\affc.\\}
\end{center}
\bigskip
\begin{abstract}
In hard interactions between external particles incident on a heat-bath,
we show that large logarithms are generated when a radiated or absorbed
gauge boson is collinear with the initial fermion momentum. These
logarithms can be absorbed into process independent splitting/absorption
probabilities. Unlike the zero-temperature case, however, they depend
explicitly on the temperature and the scale of the interaction.
\end{abstract}
\end{titlepage}
\setcounter{footnote}{0}

In recent years the problem of calculating scattering cross sections
or rates for external particles travelling in a thermal medium has
become phenomenologically interesting \cite{pheno}. In a previous paper
we have shown that to one-loop order there are no infrared divergences
when two external particles collide in a heatbath kept at a temperature
$T$ \cite{old}. We were able to resum the one-loop result and calculate
the distribution of the mismatch in initial and final momenta due to
soft radiation. The result was finite and contained some large logarithms
of the type which usually arise from collinear emissions. In this paper
we study this collinear behaviour in detail. We find that divergences
exist, and are signalled by the familiar ``large terms'' of the order
$\log(Q^2/m^2)$ ($Q^2$ is the scale of the process and $m$ is a regulating
mass). We show that the collinear part gives rise to a certain universal
splitting/absorption probability which explicitly depends on $Q$ and $T$.

There has been recent interest in collinear singularities associated with
soft particles \cite{many}, because they might spoil the hard thermal loop
resummation \cite{htl}. Since we study a hard particle in a heat-bath, we
have nothing further to say about this.

Consider the simplest scattering process in which the collinear
singularities make their appearance. This is the inclusive cross section
for the scattering of a charged electron with a space-like photon,
$\gamma^*$, in a QED heat-bath, which we consider in a perturbation theory
of the gauge coupling. The initial electron momentum is denoted
by $p$, the $\gamma^*$ momentum by $q$ and the final electron momentum
by $p'$. The 4-velocity of the heat-bath in any frame, $u$ ($u^2=1$),
is a new vector in the problem \cite{wel82}. Compared to the $T=0$ case,
 there are extra scalars $p\cdot u$ and $q\cdot u$ which have to be taken
into account. For the rest, we use the standard notation
\begin{equation}
   q^2=-Q^2\qquad{\rm and}\qquad p\cdot q={Q^2\over2x},
\label{int:kine}\end{equation}
and work in the limit $Q^2\gg m^2$, where $p^2=p'^2=m^2$. We also take
$Q^2\gg T^2$ where $T$ is the temperature of the heat-bath.

The inclusive cross section can be written as a perturbation expansion in
the form
\begin{equation}
   \sigma=\alpha\left(\sigma_0+\alpha\sigma_1+\alpha^2\sigma_2+\cdots\right),
\label{int:cross}\end{equation}
where $\alpha$ is the gauge coupling. At the lowest order in perturbation
theory, one has only the interaction between $e$ and $\gamma^*$, and hence
$\sigma_0$ is independent of $T$. At higher orders, interactions with thermal
photons have to be taken into account, and $\sigma_i$ ($i>0$) may change
from its $T=0$ value.

It is useful to write the cross section in the form
\begin{equation}
   \sigma\;=\;{\rho^{\mu\nu} W_{\mu\nu}\over4\sqrt{m^2Q^2+(p\cdot q)^2}},
                   \qquad{\rm where}\qquad
   \rho_{\mu\nu}=\sum_\lambda \epsilon^{*\lambda}_\mu(q)
          \epsilon^\lambda_\nu(q).
\label{int:sfdef}\end{equation}
$\epsilon^\lambda_\mu(q)$ is the polarisation vector of the off-shell
photon in the polarisation state $\lambda$. The tensor $\rho$ is a density
matrix for the polarisation states of $\gamma^*$ and is symmetric in its
indices. The quantity to be computed in perturbation theory is the symmetric
part of the rank-2 tensor $W_{\mu\nu}$, which is the vacuum expectation
value of the product of the electromagnetic current coupling to $\gamma^*$.

The computation is simplified by decomposing the tensor $W_{\mu\nu}$ into
scalar functions multiplying all symmetric tensors which can be built out
of the vectors in the problem. Furthermore, the gauge invariance of the
current implies that only those tensors orthogonal to $q$ are relevant.
There are four such tensors---
\begin{equation}\begin{array}{rl}
   T^1_{\mu\nu}\;=\;g_{\mu\nu}
        +{\displaystyle 1\over\displaystyle Q^2}q_\mu q_\nu\quad&\quad
   T^2_{\mu\nu}\;=\;P_\mu P_\nu\\
   T^3_{\mu\nu}\;=\;U_\mu U_\nu\qquad\quad\;\quad&\quad
   T^4_{\mu\nu}\;=\;U_\mu P_\nu+U_\nu P_\mu.\\
\end{array}\label{apten:basis}\end{equation}
We have used a shorthand notation for the components of $p$ and $u$
orthogonal to $q$---
\begin{equation}
   P_\mu\;=\;p_\mu+{\displaystyle p\cdot q\over \displaystyle Q^2}q_\mu,
   \quad{\rm and}\quad
   U_\mu\;=\;u_\mu+{\displaystyle u\cdot q\over \displaystyle Q^2}q_\mu.
\label{apten:ortho}\end{equation}

As a result,
\begin{equation}
   W_{\mu\nu}=\sum_{i=1}^4 W_i(x,Q^2,p\cdot u,q\cdot u) T^i_{\mu\nu},
\label{apten:sfdef}\end{equation}
and hence there are four ``structure functions'' in this problem. At
$T=0$ only the two structure functions $W_1$ and $W_2$ appear. Even for
$T>0$, at the leading order of perturbation theory $W_3=W_4=0$, since
$\sigma_0$ does not contain any terms involving $u$. Hence the Callan-Gross
relation \cite{cg} is also valid to this order, with corrections generated
at higher orders, through the usual vacuum ($T=0$) processes, as well as
by additional interactions with real thermal gauge bosons.

\begin{figure}
\vskip9truecm
\includegraphics{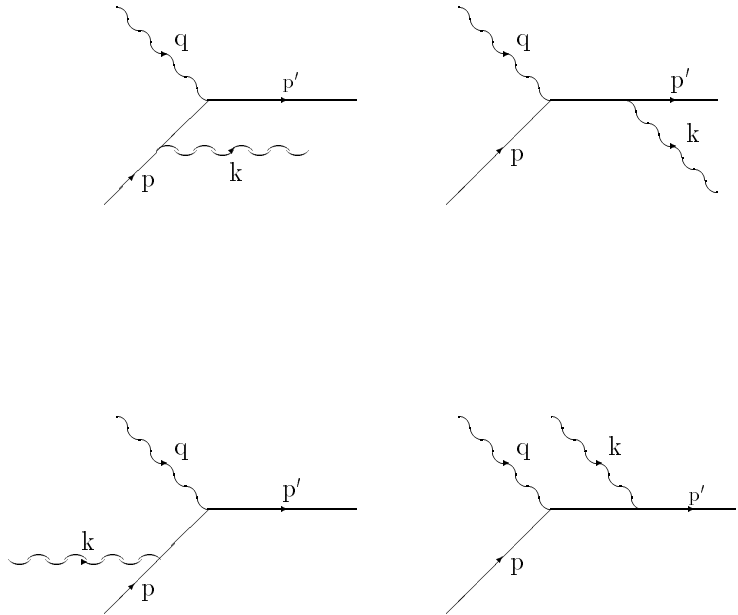}
\caption[dummy]{The real photon emission and absorption corrections.
   Absorbed thermal photons come in from the left, emitted photons exit
   to the right. In the planar gauge only the diagrams with photons
   attached to the initial leg contribute to the collinear limit.}
\label{fg:feynr}\end{figure}

Reviews of the real-time thermal field theory techniques we use can be found
in \cite{realtime}. Using these we can generate the processes contributing
to $\sigma_1$ and the rules for their evaluation. The relevant diagrams
with real thermal photon emission and absorption are shown in Figure
(\ref{fg:feynr}). For one photon emission (absorption) the two-body phase
space measure can be taken to be
\begin{equation}
   d\Gamma_\pm\;=\;
     {1\over(2\pi)^4}d^4k 2\pi\delta^\pm(k^2)2\pi\delta^+((p+q-k)^2)
       B(k\cdot u),
\label{apsud:measure}\end{equation}
where $k$ is the four-momentum of the thermal photon, $\delta^\pm(x^2)=
\delta(x^2)\theta(\pm x_0)$ and $B(x)$ is the Bose distribution
$1/[\exp(|x|/T)-1]$. The positive sign in eq.~(\ref{apsud:measure})
corresponds to the emission process and negative to absorption.

Note that for the diagrams in Figure (\ref{fg:feynr}), the vector $u$
appears only in the measure. This leads to a singularity as $k\cdot u\to0$.
However, this is not a collinear singularity but the previously analysed
soft singularity \cite{old}. Since $u^2=1$, a boost to the rest frame of
$u$ can always be done. This gives us the correct interpretation of the
divergence. The only collinear singularities then arise from the matrix
elements. These can be identified as divergences in the limit $k\cdot p\to0$
when $m\to0$, and have the same origin as those occurring at $T=0$.

We choose to work in a planar gauge \cite{ddt}, which is a ghost-free gauge
specified by the gauge fixing part of the Lagrangian---
\begin{equation}
    {\cal L}_{gf}\;=\; -{1\over2 v^2}(v_\mu A^\mu)\partial^2(v_\nu A^\nu).
\label{pert:gauge}\end{equation}
The vector
\begin{equation}
   v_\mu\;=\;C_1 p'+ C_2 p
\label{pert:gaugedef}\end{equation}
defines the gauge choice. The coefficients $C_1$ and $C_2$ are chosen
such that $v^2\ne0$ and the sum over polarisations of the real photons
becomes
\begin{equation}
   d_{\alpha\beta}\;\equiv\;
   \sum_\lambda\epsilon_\alpha^\lambda(k)\epsilon_\beta^{\lambda *}(k)
     \;=\;-g_{\alpha\beta}+{\displaystyle k_\alpha v_\beta
                +k_\beta v_\alpha\over k\cdot v},
\label{pert:completeness}\end{equation}
where $\epsilon_\mu^\lambda(k)$ is the polarisation vector of a photon
with polarisation $\lambda$ and momentum $k$.

It can be verified that in this gauge all collinear singularities come
from the squares of the diagrams with the real photon attached to the
initial fermion leg. For the first emission diagram, we obtain
\begin{equation}\begin{array}{rl}
   |M|^2\;=&\;-2e^4{\displaystyle1\over\displaystyle(p-k)^2}\\&{\rm Tr}
    \left[\gamma_\nu p'\!\!\!/ \gamma_\mu \left\{
      p\!\!/\left({\displaystyle2p\cdot v\over\displaystyle k\cdot v}-1\right)
     +k\!\!/\left(1-{\displaystyle p\cdot v\over\displaystyle k\cdot v}\right)
     +v\!\!/{\displaystyle k\cdot p\over\displaystyle k\cdot v}\right\}\right].
\end{array}\label{coll:eq2}\end{equation}

We are interested in extracting the leading terms in the collinear limit,
$p\cdot k\to0$. The most transparent way of doing this is to use the Sudakov
parametrisation \cite{ddt}---
\begin{equation}
   k\;=\;(1-\rho)p+\beta(q+xp)+k_{\scriptscriptstyle T},
         \qquad{\rm where}\qquad
    p\cdot k_{\scriptscriptstyle T}=q\cdot k_{\scriptscriptstyle T}=0.
\label{coll:sud}\end{equation}
It is clear that this is a Lorentz invariant decomposition. The integration
variables are changed to $\rho$, $\beta$ and the two independent components
of $k_{\scriptscriptstyle T}$. The Jacobian is simply
\begin{equation}
   {d^4k\over d\rho d\beta d^2k_{\scriptscriptstyle T}}\;=\;{Q^2\over2x}.
\label{apsud:jacob}\end{equation}
The variables $\rho$ and $\beta$ are fixed by the $\delta$-function
constraints in the measure $d\Gamma_+$. In the collinear limit, the
solution which leads to $p\cdot k\to0$ as $k_{\scriptscriptstyle T}^2\to0$ is
\begin{equation}
   \rho\;=\;x+{\cal O}\left(k_{\scriptscriptstyle T}^2\right),\qquad\beta\;=\;
      {\displaystyle xk_{\scriptscriptstyle T}^2\over\displaystyle Q^2(1-x)}
      +{\cal O}\left(k_{\scriptscriptstyle T}^4\right).
\label{apsud:coll}\end{equation}
Requiring $\rho$ and $\beta$ to be real, in the limit $m\to0$ we find
\begin{equation}
   0\;\le\;k_{\scriptscriptstyle T}^2\;\le\;{Q^2\over4x}(1-x).
\label{apsud:limits}\end{equation}
The $\theta$-functions place no further restrictions, and may be
dropped to give the phase space measure
\begin{equation}
  d\Gamma_+\;=\;
    {\displaystyle xd\rho d\beta d^2k_{\scriptscriptstyle T}
       \over\displaystyle 2Q^2(1-x)(2\pi)^2}
   \delta\left(\beta-{xz\over1-\rho}\right)\delta(\rho-x).
\label{apsud:final}\end{equation}

Absorption is handled by making the change of variables $k_\mu\to-k_\mu$,
and then writing a Sudakov parametrisation as before. $d\Gamma_-$ differs
from $d\Gamma_+$ only in the change
\begin{equation}
   \delta(\rho-x)\;\longrightarrow\;\delta(\rho-2+x).
\label{apsud:finalm}\end{equation}

In the collinear limit, the denominator on the right of eq.~(\ref{coll:eq2})
becomes zero, and hence some care is required in taking this limit. We
retain the fermion mass, $m$, as a regulator in the denominator and write
\begin{equation}
 {1\over(p-k)^2}\;=\;{1\over m^2(1-2\rho)-Q^2\beta/x}.
\label{coll:den}\end{equation}
The most singular terms in the collinear limit can then be found by
evaluating the trace in eq.~(\ref{coll:eq2}) for $m=0$ and neglecting terms
in $k_{\scriptscriptstyle T}^2$ (and hence in $\beta$). After some
straightforward manipulations, we find that the most singular contribution
to the tensor $W_{\mu\nu}$ is
\begin{equation}\begin{array}{rl}
   -8e^4&\left[2xT^2_{\mu\nu}
           -{\displaystyle Q^2\over\displaystyle2x}T^1_{\mu\nu}\right]\\
     &\;\;\int d\Gamma_+\,B[(1-\rho)p\cdot u]
    \left({\displaystyle1+\rho^2\over\displaystyle1-\rho}\right)
            {\displaystyle1\over\displaystyle m^2(1-2\rho)-Q^2\beta/x}
\end{array}\label{coll:eq3}\end{equation}
The contribution from the corresponding absorption diagram, obtained
by changing $k\to-k$ in the matrix element and using the phase space
measure $d\Gamma_-$, is simply obtained by setting $\rho$ to $2-\rho$
is the above expression.

In the chosen gauge, the squares of the other two diagrams do not have
singular denominators and hence can be neglected in the collinear limit.
The cross terms do have a singular denominator. However, the trace contains
\begin{equation}
   \stackrel{\displaystyle{\rm Lt}}{\scriptscriptstyle k\to(1-\rho)p}
               d_{\alpha\beta}(k)p^\alpha\;=\;0.
\end{equation}
Hence this term can also be neglected.

Two facts about eq.~(\ref{coll:eq3}) and its analogue for photon absorption
are worth pointing out. First, both terms contain only the tensors $T^1$ and
$T^2$. Hence there are no collinear contributions to the thermal structure
functions $W_3$ and $W_4$ at this order.
Second, the integrand has a probability interpretation. The part
$(1+\rho^2)/(1-\rho)$ can be interpreted as the probability that
an incoming electron radiates a photon carrying a fraction $1-\rho$ of the
initial momentum. The Bose distribution factor is the probability that
the radiated photon is indistinguishable from a thermal photon. The
integrand has support on $0\le\rho\le1$, which is consistent with this
interpretation. Similarly, the absorption process can also be interpreted
as the product of two probabilities. The Bose distribution is the probability
of finding a photon in the heat-bath carrying a fraction $\rho-1$ of the
incoming electron's momentum and the factor $(1+(2-\rho)^2)/(1-(2-\rho))$
is the probability of absorption. This integral has support on $1\le\rho\le2$.

The integrals can be performed completely. The thermal leading log part is
\begin{equation}
   W_{\mu\nu}\;=\;4\pi\alpha^2 P(x,Q/T)
       \log\left({\displaystyle Q^2\over \displaystyle m^2}\right)
      {\displaystyle x\over\displaystyle Q^2}
     \left[2xT^2_{\mu\nu}
           -{\displaystyle Q^2\over\displaystyle2x}T^1_{\mu\nu}\right].
\label{coll:result}\end{equation}
We have introduced the finite temperature part of the ``splitting function''
\begin{equation}
     P(x,Q/T)\;=\;{2\over\exp[(1-x)p\cdot u/T]-1}\,
            \left({\displaystyle 1+x^2\over\displaystyle1-x}\right).
\label{coll:splitting}\end{equation}
This can be given an interpretation as the probability of an external
electron splitting off a thermal photon. Unlike the case at $T=0$, this
factor has an explicit dependence on $Q/T$. Note also that $P(x,Q/T)$ has
a singularity as $x\to1$. This is the region of phase space where the
thermal photon is soft. We have shown in \cite{old} that the cross section
is finite in this limit provided virtual corrections are taken into
account. Using this result, we can simply write down a regulated version
of eq.~(\ref{coll:splitting}) as the splitting probability for the external
electron.

At $T=0$ the leading soft divergence in the real diagrams is logarithmic,
and shows up in the splitting functions as a divergence of the form
$1/(1-x)$, in the limit $x\to1$. It is cured by taking into account the
virtual diagrams. The regulated form of the splitting functions is then
given by the familiar prescription
\begin{equation} 
   \int dx P_+(x) f(x)\;=\;\int dx P(x)\left[f(x)-f(1)\right],
     \qquad\qquad(T=0),
\label{appl:zero}\end{equation}
where $f(x)$ is a test function. Note that the first moment of any
distribution vanishes when convoluted with the splitting function
so regularised.

For $T>0$ the leading soft divergence is quadratic. This is signalled by
a divergence of the form $1/(1-x)^2$ ($x\to1$) in the splitting functions.
Its cancellation against virtual contributions has been shown in \cite{old}.
The sub-leading logarithmic divergence has also been shown to cancel
against virtual corrections \cite{indu}. Consequently, we just write down
an appropriately regularised version of the $T>0$ part of the splitting
function---
\begin{equation} 
   \int dx P_+(x) f(x)\;=\;\int dx P(x)
      \left[f(x)-f(1)-(x-1) f'(1) \right].
\label{appl:def}\end{equation}
Note that the first two moments of $P_+$ vanish with this regularisation.
The vanishing second moment implies that finite temperature effects do not
change the expectation value of the parton's momentum. This is expected
\cite{old} and is a consequence of detailed balance.

The universality of this additional finite temperature term in the
splitting functions derived here can be easily checked. The calculation
for Fermion pair-production is very similiar to the computation
presented in this paper, and yields precisely the same collinear
term derived here.

In QED, since the electron mass is non-zero, our results are complete.
However, in a real experiment we shall have to deal with a QCD heatbath.
Our main results, eq.~(\ref{coll:result}), along with
eqs.~(\ref{coll:splitting}) and (\ref{appl:def}), can be carried over
to this case with the simple replacement
$\alpha^2\to C_{\scriptscriptstyle F}\alpha\alpha_{\scriptscriptstyle S}$.
The crucial change is that $m=0$ for quarks and hence the results are
singular.

At $T=0$, these collinear singularities are handled by factoring them
into universal quark distributions inside hadrons. A similiar procedure
will have to be developed for external hadrons or jets impinging on a
plasma. In order to complete this program, a suitable definition of the
QCD running coupling at finite temperature \cite{baier} must be provided.
At $T=0$, this is sufficient information to sum the one-loop iterated
ladder diagrams into the DGLAP equations \cite{dglap}.

For $T>0$ the situation is more complicated. This is clear from the fact
that the analogue of the splitting function contains the dimensionless
variable $Q/T$ in addition to $x$. At finite temperature and arbitrary
scale $Q^2$, we are forced to consider two scales in the renormalisation
group \cite{twoscale}. In various domains these simplify. For example,
when $Q\gg T$, one expects to be able to use a single scale. In this
limit, the regularised version of eq~(\ref{coll:splitting}) vanishes,
and the evolution in $Q^2$ is the same as at $T=0$. However, the parton
distributions at each $T$ must then be seperately measured. Only by
keeping two scales can information at $T=0$ be evolved to $T>0$.
This work is left to the future.


\begin{thebibliography}{99}
\bibitem{pheno}
   M.\ Gyulassy and M.\ Pl\"umer, {\sl Nucl.\ Phys.\/} B 346 (1990) 1;\\
   M.\ Gyulassy and X.-N.\ Wang, {\sl Nucl.\ Phys.\/} B 420 (1994) 583;\\
   J.-C.\ Pan and C.\ Gale, {\sl Phys.\ Rev.\/}, D 50 (1994) 3235;\\
   S.\ Gupta, {\sl Phys.\ Lett.\/}, B 347 (1995) 381.
\bibitem{old}
   S.\ Gupta, D.\ Indumathi, P.\ Mathews and V.\ Ravindran,
      {\sl Nucl.\ Phys.\/} B 458 (1996) 189;\\
   H.\ A.\ Weldon, {\sl Phys.\ Rev.\/}, D 49 (1994) 1579.
\bibitem{many}
   R.\ Baier, S.\ Peigne and D.\ Schiff, {\sl Z.\ Phys.\/}, C 62 (1994) 337;\\
   F.\ Flechsig and A.\ K.\ Rebhan, {\sl Nucl.\ Phys.\/}, B 464 (1996) 279;\\
   P.\ Aurenche, F.\ Gelis, R.\ Kobes and E.\ Petitgirard, hep-ph/9604398
       (to appear in {\sl Phys.\ Rev.\/}, D), and hep-ph/9609256.
\bibitem{htl}
   R.\ D.\ Pisarski, {\sl Phys.\ Rev.\ Lett.\/}, 63 (1989) 1129;\\
   E.\ Braaten and R.\ D.\ Pisarski, {\sl Nucl.\ Phys.\/}, B 337 (1990) 569;\\
   J.\ Frenkel and J.\ C.\ Taylor, {\sl Nucl.\ Phys.\/}, B 334 (1990) 199.
\bibitem{wel82}
   H.\ A.\ Weldon, {\sl Phys.\ Rev.\/}, D 26 (1982) 1394.
\bibitem{cg}
   C.\ G.\ Callan D.\ Gross, {\sl Phys.\ Rev.\ Lett.\/}, 22 (1969) 156.
\bibitem{realtime}
   N.\ P.\ Landsman and Ch.\ G.\ van Weert,
       {\sl Phys.\ Rep.\/}, 145 (1987) 141;\\
   A.\ Niemi and G.\ Semenoff, {\sl Nucl.\ Phys.\/}, B 230 (1984) 181.
\bibitem{ddt}
   Dokshitzer, Yu.\ L.\, Dyakonov, D.\ I.\ and Troyan, S.\ I.,
   {\sl Phys.\ Rep.\/} 58 (1980) 269.
\bibitem{indu}
   D.\ Indumathi, Dortmund preprint DO-TH-96-09, hep-ph/9607206.
\bibitem{baier}
   R.\ Baier, B.\ Pire and D.\ Schiff, {\sl Phys.\ Lett.\/}, B 238 (1990) 367.
\bibitem{dglap}
   G.\ Altarelli and G.\ Parisi, {\sl Nucl.\ Phys.\/}, B 126 (1977) 298;\\
   V.\ N.\ Gribov and L.\ N.\ Lipatov, {\sl Sov.\ J.\ Nucl.\ Phys.\/},
     46 (1972) 438.
\bibitem{twoscale}
   J.\ Kapusta, {\sl Finite Temperature Field Theory\/}, Cambridge
     University Press, Cambridge, UK, 1989;\\
   C.\ Ford and C.\ Wiesendanger, preprint DIAS-STP 96-10, hep-ph/9604392.
\end{thebibliography}
\end{document}